\documentclass[preprint2,twoside]{hwo}

\usepackage{lipsum}
\usepackage{natbib}

\newcommand{\msun}{{M$_\odot$}}
\newcommand{\zsun}{{Z$_\odot$}}
\newcommand{\heiiuv}{HeII~1640}
\newcommand{\heiiopt}{HeII~4686}
\newcommand{\nivuva}{NIV~1486}
\newcommand{\nivuvb}{NIV~1720}
\newcommand{\civopt}{CIV~5801-12}

\DeclareUnicodeCharacter{2212}{-}

\input{hwo.h}

\begin{document}

\title{\textbf{\LARGE Very Massive Stars with the Habitable Worlds Observatory}}
\author {\textbf{\large Fabrice Martins,$^{1}$ Aida Wofford,$^2$ Miriam Garcia,$^3$ Peter Senchyna,$^4$ Janice Lee,$^{5,6}$ Paul A. Scowen$^7$}}
\affil{$^1$\small\it LUPM, Univ. Montpellier, CNRS, Montpellier, Place Eugene Bataillon, F-34095, France}
\affil{$^2$\small\it Instituto de Astronom\'ia, Universidad Nacional Aut\'onoma de M\'exico, Unidad Académica en Ensenada, Km 103 Carr. Tijuana−Ensenada, Ensenada, B.C., C.P. 22860, M\'exico}
\affil{$^3$\small\it Centro de Astrobiolog\'ia, CSIC-INTA. Crtra. de Torrej\'on a Ajalvir km 4., E-28850 Torrej\'on de Ardoz (Madrid), Spain}
\affil{$^4$\small\it The Observatories of the Carnegie Institution for Science, 813 Santa Barbara Street, Pasadena, CA 91101, USA}
\affil{$^5$\small\it Space Telescope Science Institute, 3700 San Martin Drive, Baltimore, MD 21218, USA}
\affil{$^6$\small\it Steward Observatory, University of Arizona, 933 North Cherry Avenue, Tucson, AZ 85721, USA}
\affil{$^7$\small\it NASA/GSFC, Mail Code 667, Greenbelt, MD 20771}

\author{\footnotesize{\bf Endorsed by:}
St\'ephane Blondin (CNRS, Aix-Marseille University), Jean-Claude Bouret (CNRS, Aix-Marseille University), Paul Crowther (University of Sheffield), Alexandre David-Uraz (Central Michigan University), Ruben Joaquin Diaz (Noirlab), Jose M. Diego (Instituto de Fisica de Cantabria), Christi Erba (Space Telescope Science Institute), Sophia Flury (University of Edinburgh), Luca Fossati (Space Researh Institute, Austrian Academy of Sciences), Kevin France (University of Colorado), Alex Fullerton (Space Telescope Science Institute), Joris Josiek (Heidelberg University, ZAH/ARI), Pierre Kervella (Paris Observatory, CNRS IRL FCLA), Gloria Koenigsberger (UNAM Instituto de Ciencas Fisicas), Brad Koplitz (Arizona State University), Ji\u r\'i Krti\u cka (Masaryk University), Iva Krti\u ckov\'a (Masaryk University), Ariane Lan\c{c}on (Astronomical Observatory of Strasbourg), Eunjeong Lee (Eiskosmos, CROASAEN), Roel Lefever (Heidelberg University, ZAH/ARI), Siliva Martocchia (CNRS, Aix-Marseille University), Donatas Narbutis (Vilnius University), Gioia Rau (NSF), Andreas Sander (Heidelberg University, ZAH/ARI), Linda Smith (Space Telescope Science Institute), Frank Soboczenski (University of York \ King's College London), Heloise Stevance (University of Oxford), Robert Szabo (Konkoly Observatory), Grace Telford (Princeton University), Asif ud-Doula (Penn State University), Ankur Upadhyaya (University of Warwick)
}

\begin{abstract}
  Very massive stars (VMS) are defined as stars with an initial mass in excess of 100~\msun. Because of their short lifetime and the shape of the stellar mass function, they are rare objects. Only about twenty of them are known in the Galaxy and the Large Magellanic Cloud. However VMS are important in several ways. They efficiently spread nucleosynthesis products through their boosted stellar winds, they are predicted to explode as pair-instability supernovae or to form heavy black-holes from direct collapse, and they outshine all other types of stars in the ultraviolet light, thus dominating the integrated light of starbursts. Their presence is indirectly suspected across all redshifts, all the way to cosmic dawn where they may have played a key role in the formation of the first galaxies. Their search and identification is currently hampered by instrumental limitation, especially spatial resolution. An integral field spectrograph working at the diffraction limit of HWO (5~mas) and with a spectral resolution of about 2000 would revolutionize the understanding of VMS. We make the case for such an instrument in this contribution.
  \\
  \\
\end{abstract}

\vspace{2cm}

\section{Science goal: what are the most massive stars?}

When looking at stars differences can be seen in their brightness and their color. The former is related to the ability of stars to produce energy and emit it in the form of light. The latter is rooted in the temperature at their surface. For most of the stars we observe in the sky the energy production and the temperature are related to a third property: the mass of the star. On average, more massive stars are more luminous and hotter. Some deviations from this trend exist in specific phases of the evolution of stars but to first order this is the rule. Mass is thus the most fundamental parameter that sets the appearance of a star.

The distribution of mass among stars is not random. If we take the Sun as a reference (its mass, noted \msun, is about $2 \times 10^{30}$ kg), stars that have a mass below that of the Sun are more numerous than those that have a mass above it. Conversely stars more massive than the Sun are less numerous: one typically finds two hundred less stars with a mass of ten times that of the Sun compared to Sun-like stars. The lower limit a star can have is of the order seven hundredths of the Sun’s mass. Below that limit no energy production (through nuclear reaction) is possible and the object is not a star, by definition. 

On the other side of the mass distribution, the maximum mass a star can reach is unknown. One of the reasons is the small number of high-mass stars compared to lower mass stars, as described above. Finding high-mass stars is thus more difficult. The other main reason is that a massive star lives much shorter than a lower mass star. The lifetime of the Sun is about 10 billions of years, while that of a 10~\msun\ star is only 100 millions of years. Consequently spotting massive stars among other stars is difficult not only because there are few of them but also because they are present for a much shorter time than lower mass stars.

As of today the most massive star we know has a mass of about 200 times that of the Sun \citep{brands22}. Above 100 \msun\ only ten to twenty stars have been firmly identified. These objects are known as Very Massive Stars or VMS. The present science case aims at describing how HWO will discover more of these objects.

Since VMS are extremely rare compared to other stars, one may wonder whether finding them is important. Massive stars in general, and VMS in particular, have many unique properties. They produce most of the elements from oxygen to iron and even more complex atoms. The elements that make most of the Earth’s composition (oxygen and silicon in the crust, and iron in the core) have thus been produced by massive stars before the formation of the Earth. These elements have been expelled in the stars’ surroundings by the strong mass ejections they experience, either in the form of winds  or at the end of their lives when they explode as supernovae. In addition to chemical elements mechanical energy is also released through winds and explosions, carving bubbles in the gas surrounding massive stars and triggering new episodes of star formation. The leftovers of massive star evolution are neutron stars or black holes. Some of them can merge and drive gravitational waves that are detected on Earth. 

VMS have additional properties. Their winds are much stronger than those of normal massive stars (e.g. \citealt{besten20b,graef21}), which boosts their ability to inject freshly synthesized elements in the close environment. VMS are also extremely luminous. This compensates for their very small number: a few VMS can completely dominate the total light of a cluster that also contains thousands of less massive stars \citep{crowther16,crowther24}. This property is particularly attractive to explain one of the most intriguing results of the James Webb Space Telescope (JWST) observations, namely the overabundance of massive galaxies in the early Universe (see \citealt{stark25} for a review). Since the mass of galaxies is determined from the amount of light they emit, accounting for the presence of VMS can reduce the estimated mass of the far-away galaxies and relieve part of the tension with models of galaxy formation \citep{schaerer25}. JWST has also shown that some of these galaxies contain much more nitrogen than expected \citep{bunker23,marques24,naidu25}. Some of these galaxies may be the precursors of globular clusters which are old clusters made of stars with puzzling chemical composition \citep{charbonnel23,senchyna24}. The formation of globular clusters, that exist in many galaxies, is not understood at present. Because of their efficient production of elements, VMS are serious candidates to explain the chemical peculiarities of the stars nested in globular clusters \citep{vink23}. VMS are thus key players of galaxy growth (one of the three NASA decadal priorities), especially in the early Universe.

The current number of known VMS is restricted by observational limitations. Since VMS live short and exist in very small numbers, one has to probe large samples of stars to spot them. This means that on average one has to look at large distances to be lucky enough to find them. But as distance increases, confusion between stars increases because of the limitation of the current generation of telescopes to separate objects from each other. This is where HWO will make a difference: with its preferred position in space, its large mirror and its ability to observe in the ultraviolet and optical, it will uncover VMS in new environments and revolutionize our understanding of the formation, evolution and fate of VMS. Only HWO will be able to achieve the image sharpness in the optical and UV ranges needed to characterize these objects. ELTs may be able to isolate them in other ranges of the electromagnetic spectrum, but the key diagnostics needed to understand VMSs are only within HWO’s grasp.

\section{Science Objective}

\subsection{Uncovering and characterizing individual Very Massive Stars in the Local Universe with spatially resolved UV-optical spectroscopy}

\begin{figure}[!ht]
\includegraphics[width=\columnwidth]{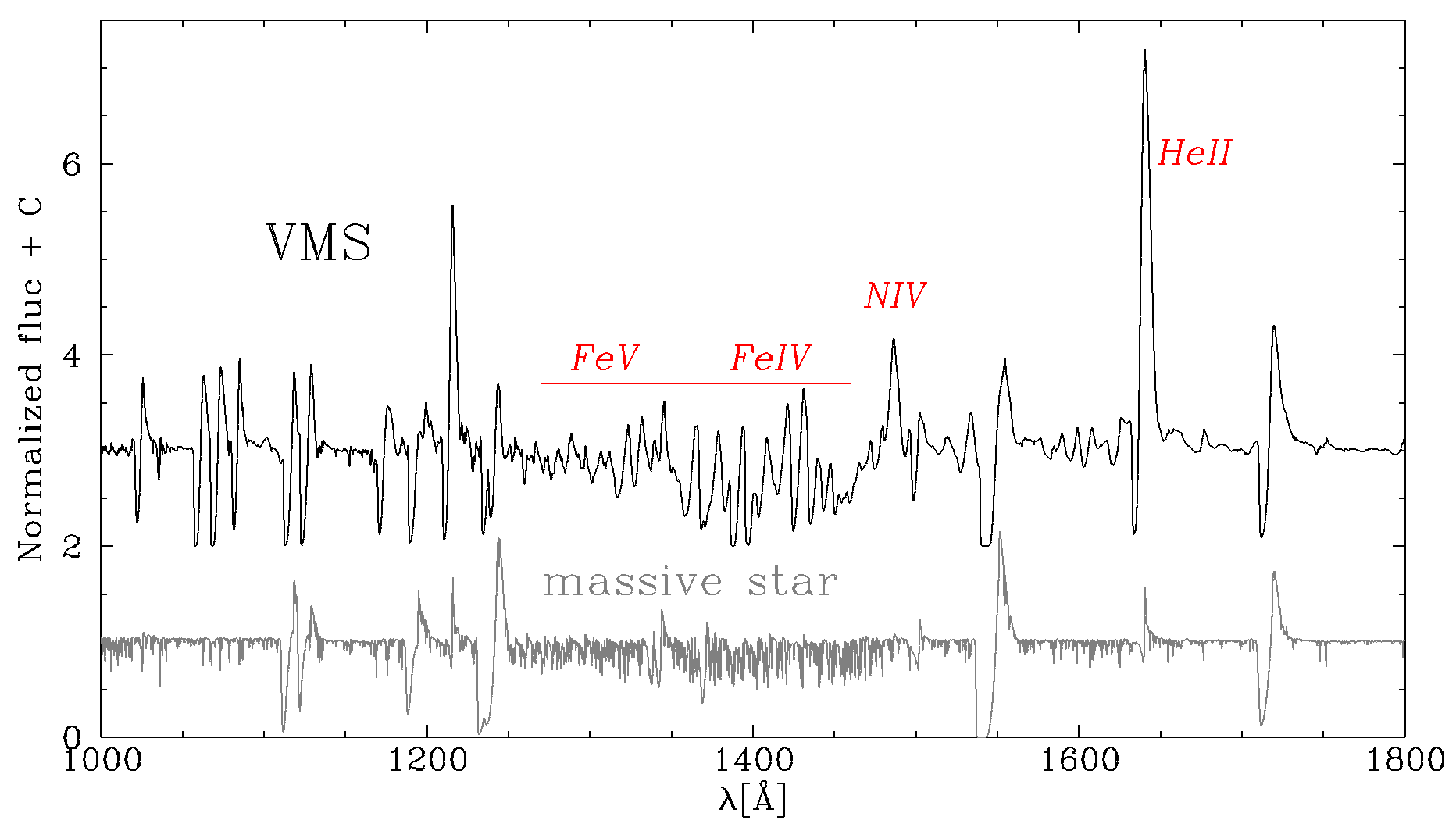}
\caption{\small Theoretical UV spectrum of a VMS (top black line) compared to a normal massive star’s spectrum (bottom grey line). The VMS spectrum is dominated by a strong \heiiuv\ emission. \nivuva\ emission is seen in VMS but not in normal O-type stars. Spectra are from \citet{mp21}.}
\label{fig_vms_normal}
\end{figure}

VMSs are defined as stars with a mass in excess of 100~\msun\ \citep{vink15}. Compared to normal massive stars they are mostly characterized by a higher luminosity, boosted mass loss rates, and potentially different end points. Because of the relation between mass and luminosity VMS have luminosities higher than 10$^6$ times the luminosity of the sun (Lsun). At the same time they are hot (Teff of the order 40000~K and above) which implies that their spectral energy distribution peaks in the far-UV, that VMS emit copious amounts of ionizing photons and that their light dominates that of all other stellar sources in this wavelength range. \citet{crowther16} showed that the seven most massive stars of the Large Magellanic Cloud star cluster R136 contribute 30\% of the FUV continuum light of the entire cluster, and are the only contributors to the \heiiuv\ UV emission line (see Fig.~\ref{fig_vms_normal}). As described below this feature is a powerful diagnostic feature of VMS. Because of their high luminosity VMS experience mass loss rates that are not just scaled-up versions of normal massive stars. Their winds are about ten times stronger than would be expected from simple extrapolation of relations between mass loss rates and luminosity for normal massive stars \citep{martins08,gh08,besten20a}. This strongly affects both the evolution and the spectral appearance of VMS. In particular the \heiiuv\ line mentioned above directly results from these effects. The end points of the evolution of VMS are relatively unconstrained. VMS, at least some of them, are suspected to go through pair-instability supernovae and leave no remnants behind them. The most massive ones may skip that phase and form heavy black holes that can serve as seeds for the formation of supermassive black holes at the heart of galaxies.

As of today very few VMS have been identified. The best laboratory has so far been the core of the giant HII region 30 Dor in the Large Magellanic Cloud, i.e. the cluster R136. Here 5 to 10 VMS have been discovered. The exact number depends on the mass estimate which itself relies on evolutionary tracks and luminosity uncertainties. Among them four stars reach 150 to 300 Msun (stars R136 a1, a2, a3 and c – see \citealt{crowther10,brands22}). Three other VMS lie at the core of the young cluster NGC3603 in the Milky Way \citep{cd98,crowther10}. They have masses between 100 and 170 Msun. Near the Galactic Center the Arches cluster host a handful of stars with masses just above 100 Msun \citep{martins08}. \citet{bru03} report the presence of two stars with high enough luminosity to be classified as VMS in the giant HII region NGC604 of M33. Finally a few individual stars located in star forming regions may be classified as VMS based on their high luminosity \citep{figer98,hamann06,barniske08,besten11}. In total we therefore know about 20 VMS with only about five of them having masses clearly above 150~\msun. 

Because of their strong winds and rapid evolution VMS show specific spectral features in the UV and optical ranges. The fast evolution of VMS triggers efficient chemical mixing that enriches their surfaces \citep{yusof13,kohler15,graef21,mp22,sab22,martinet23,kes25}. The products of hydrogen burning through the CNO cycle are visible, namely a strong nitrogen and helium enrichment. Coupled to a strong wind, with boosted mass loss rates, this explains the spectroscopic appearance depicted in Fig.~\ref{fig_vms_normal}. In the ultraviolet, \heiiuv\ and \nivuva\ show strong emission and/or P-Cygni profiles that are either not seen or much weaker in normal massive stars. In the optical, VMS are dominated by \heiiopt\ emission and a narrow emission from the \civopt\ doublet \citep{martins23}. These unique features, particularly in the UV range, are thus excellent diagnostics of VMS. 

VMS are difficult to observe because of their small number and their short lifetimes. The shape of the initial mass function implies a deficiency of VMS compared to other normal (massive) stars, but also shows that VMS can only be found in star forming regions large enough to host a rich population of stars. In addition VMS live short, typically 2-3 Myr \citep{yusof13,kohler15,graef21,mp22}. We thus expect VMS to be present in intense and young star forming regions such as the clusters R136 and NGC3603 described above.

The prime goal of the present science case is to identify and characterize VMS in such clusters that remain currently unresolved, but in which the presence of VMS is suspected. The reason is that, as described above, VMS dominate the UV spectrum of clusters they live in, producing unique features such as \heiiuv\ emission. The similarity between the UV spectrum of the most massive stars in R136 and the integrated light of unresolved clusters has been raised for several clusters: NGC3125-A1 \citep{wofford14,wofford21}, NGC5253-5 \citep{smith16}, Mrk71-A \citep{smith23}, IIZw40-N \citep{leitherer18}. Studies of starbursts galaxies in the Local Universe have also highlighted the likely presence of VMS based on the morphology of UV spectra \citep{senchyna19,senchyna21,upad24,marques24}. \citet{mp22} showed that consistently taking into account VMS in population synthesis models of young starbursts was needed to reproduce the observed UV spectrum of R136 and NGC3125-A1. Models limited to stars less massive than 100 Msun never produce the \heiiuv\ line and underestimate other features (see also \citealt{wofford23}). Building on these spectroscopic properties, \citet{martins23} defined empirical criteria to select unresolved clusters likely hosting VMS. Using R136 as a template (for which the individual components are known, and the total integrated light has been obtained by HST) they set a limit of 3~\AA\ on the equivalent width of \heiiuv\ for a cluster to host VMS. Additional criteria rely on the morphology of the \heiiopt\ and \civopt\ optical features. The NIII emission blueward of \heiiopt, often seen in evolved massive stars (i.e. Wolf-Rayet stars - WR) is much weaker than \heiiopt\ in VMS. The CIV feature shows a broad emission in WR stars, but a clear doublet in emission in VMS. These features and their morphology are the key diagnostics that currently allow to distinguish VMS from more classical WR stars. They are thus the prime targets of future spectroscopic observations aiming at detecting VMS in various environments. 

The strategy to identify VMS with HWO will thus be 1) pre-select candidate clusters from the morphology of their UV and optical integrated spectra, then 2) obtain spatially resolved spectroscopy of their components to distinguish VMS from the bulk of other massive stars. The second point can be done using different observing strategies as will be described below.

\subsection{Mass loss rate of VMS as a function of metallicity}

One of the main characteristics of VMS is their strong mass loss rate. As already stated this drives their evolution and spectroscopic appearance because these rates are not just scaled-up versions of those of normal massive stars, they are boosted. The physical reason is thought to be the proximity to the Eddington luminosity, which is the maximum luminosity a star can have before the outward pressure of its radiation balances or exceeds the inward pull of gravity on the star's outer layers \citep{gh08,vink11,besten20b,graef21}. At present the relation between mass loss rate and Eddington parameter (which measures the proximity to the Eddington luminosity) has been established based on the few stars known in the LMC. This recipe is currently included in evolutionary models \citep[e.g.][]{graef21} and used to predict the spectral appearance of VMS over a wide range of evolutionary status \citep{mp21}. 
For normal massive stars, mass loss rates are known to depend on metallicity, both from empirical measurements and theoretical simulations \citep{mokiem07,marcolino22,bjorklund21}. A scaling with Z$^{0.8}$ is obtained. Similarly their descendants – WR stars – also show weaker winds at lower metallicity \citep{crowther02,hainich15,sander20}. But for the specific case of VMS, the relation with metallicity is not empirically constrained. If a reduction is expected given the behavior of stellar winds for OB and WR stars, \citet{smith23} argue that VMS keep the same wind strength at low Z, based on the strength of \heiiuv\ in the spatially unresolved cluster MrK71-A likely hosting VMS. Theoretical simulations also indicate a reduced Z-scaling for WR stars when hydrogen is present at the surface, as in the winds of VMS \citep{gh08}. The first models dedicated to VMS indicate a complex interplay between mass loss, temperature and metallicity \citep{lefever25}.

\begin{figure}[ht]
\includegraphics[width=\columnwidth]{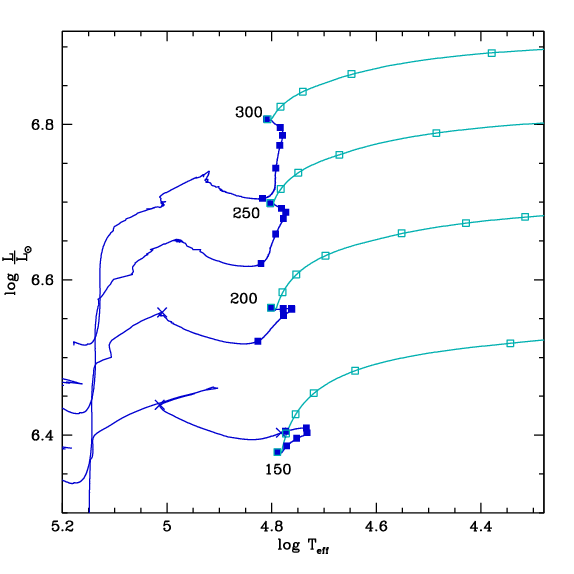}
\caption{\small Evolutionary tracks of VMS at a metallicity of 0.1 that of the Sun. The blue (cyan) lines correspond to models without (with) a metallicity scaling of mass loss rates. From \citet{martins25}.}
\label{fig_hrd}
\end{figure}

The question of the metallicity dependence of VMS winds is crucial, since pathways and final states of VMS strongly depend on their history of mass removal. Fig.~\ref{fig_hrd} illustrates this point. Evolution is drastically different depending on the assumed metallicity dependence of VMS mass loss rates. Determining the mass loss rates of a large number of VMS at metallicities lower and higher than that of the LMC is thus mandatory to make progress in the understanding of the physics of VMS. In particular the stellar yields and mechanical feedback are strongly dependent on the mass loss rates, and so are the final state of evolution. This affects the contribution of VMS to the chemistry of star-forming regions, and among others their potential role in the formation of globular clusters (see above).

\section{Physical parameters}

\subsection{Stellar parameters: how to characterize VMS}

The most common way of determining the mass of a star is to place it in the Hertzsprung-Russell (HR) diagram and compare its position to theoretical evolutionary tracks of stars with different initial masses. Consequently this requires the knowledge of two fundamental parameters: effective temperature and luminosity. 

Effective temperature is best determined by the ionization balance method that compares the strength of lines from successive ions of the same element. This is classically done in the optical range with helium and nitrogen lines \citep{massey04,martins05,rivero12,ramachandran19,martins24}. UV spectra alone can be used when no optical data are available \citep{bouret03,martins04}. In that case iron, oxygen and carbon lines are the best indicators. 

Luminosity (L) determination can be performed in two ways. The best one is to adjust the spectral energy distribution with synthetic spectra in which L can be varied and for which flux-attenuation due to dust can be applied. This method requires flux calibrated spectra and/or photometry over the widest wavelength range possible. Alternatively, photometry in a few bands (at least two for an estimate of extinction) can be used together with tabulated bolometric corrections. The latter depends on Teff that should be determined previously with the method above.

Fig.~17 of \citet{brands22} illustrates the method to estimate the mass of VMS from their Teff and L, as well as the uncertainties. The comparison of the position of stars in the HR diagram to evolutionary tracks provides the current and initial masses. Mass estimates mostly depend on the luminosity estimate, and on the luminosity predictions of models. Even if an uncertainty of 50 Msun is possible, the identification of stars clearly above 100 Msun (i.e. VMS) is robust against theoretical tracks and parameters’ determination.

The next parameters to be determined are the mass loss rate and the terminal velocity (i.e. the wind parameters). The latter is empirically obtained from the blueward extension of P-Cygni profiles \citep{leitherer92,garcia14}. Mass loss rates are determined through spectroscopic analysis with atmosphere models that adjust the wind outflow strength to reproduce the emission lines. UV spectra are most useful for that process \citep{bouret03,bouret15,hillier03,martins13,brands22} while some optical lines can also be helpful. In practice NV 1240, CIV 1550, \heiiuv, \nivuvb\ in the UV and \heiiopt\ and H$\alpha$ in the optical are good diagnostics. 

Finally surface abundances are needed to assess the evolutionary status of stars, and constrain their metallicity. Analysis of lines from CNO elements using both optical and UV lines inform on the degree of chemical mixing and the state of nucleosynthesis \citep{hunter07,hunter08,martins08,martins15,martins24,bouret13,bouret21,rivero12,besten20a,brands22}. Iron lines in the UV are sensitive to initial metallicity and can probe metallicity \citep{lanz05,bouret15,schoesser25}. Fig.~\ref{fig_vms_normal} shows that Fe lines produce intense absorption and emission lines in the far-UV that can be used for that purpose. Alternative metallicity estimates can be obtained from nebular optical lines emitted in the HII regions surrounding VMS hosting clusters. 

The determination of wind properties at different metallicities will constrain the Z scaling of mass loss rate. Constraining this relation, even partially, would constitute a major breakthrough with major consequences for the study of young starbursts in the early Universe.

\subsection{Target selection}

Candidate clusters to host VMS are pre-selected from their UV and optical spectra. As described above they should show strong \heiiuv, and potentially \nivuva, in their integrated spectrum. If only VMS, together with normal OB-type stars, are present, as is the case in R136, then \heiiopt\ should be much stronger than  the neighboring NIII~4634-41 emission. In addition \civopt\ should show a narrow doublet. Clusters showing these features are the primary targets. They will be selected using preliminary observations with existing facilities, or archival spectra. 

In some clusters there may be a mix of VMS and WR stars. The giant HII region 30 Dor is powered by NGC2070. At its center lies R136. The latter hosts only VMS, but WR stars are found in NGC2070. \citet{crowther24} provide the integrated optical and UV spectrum of NGC2070 (see their Fig.~2 and 4). The UV spectrum is still dominated by \heiiuv\ that is due to both VMS and WR stars. In the optical, \heiiopt\ remains strong compared to NIII~4634-41. But the narrow \civopt\ doublet is now diluted by the broad emission from WC stars, making it more difficult to distinguish it (although it is still present on top of the broad emission). 

VMS may thus be found in regions that also contain WR stars that dilute the VMS signatures. However a strong \heiiuv, and to a lesser extent a weak NIII 4634-41 feature relative to \heiiopt\ -- two features typical of VMS -- are still observed. Clusters showing this type of UV and optical spectra are secondary targets to search for VMS. Even if in the end VMS are not discovered in such clusters, the populations of WR stars that will be revealed are extremely interesting. Indeed as of today we don’t know of any clusters that contain normal massive and WR stars, that are fully resolved into its components, and for which the integrated optical and UV light is also available. A WR cluster template, as R136 is for VMS, does not exist. This limits our understanding of unresolved, young and massive star-forming regions. 

In a few clusters the presence of VMS has been strongly suggested \citep[see above;][]{wofford14,wofford21,leitherer18,smith16,smith23}. \citet{berg24} report the FUV spectra of about 10 clusters in M101, with \heiiuv\ seen in most of them. \citet{martins23} selected a couple of strong VMS candidates, with additional candidates. \citet{gomez21} shows the optical spectra of 38 clusters in the Antennae galaxies. More clusters in that environment are listed by e.g. \citet{whitmore10}. \citet{hadfield05} and \citet{dellabruna22} identified \heiiopt\ emission in about 150 clusters of M83, some of which showing the optical spectra typical of VMS-dominated sources. 

To investigate metallicity effects, environments sampling a wide range of 12+log(O/H) values should be selected. At present mass loss rates of VMS have been empirically calibrated only for LMC metallicity. A few VMS exist at solar metallicity but their number (about a handful) is too small to constrain any potential metallicity variation of mass loss rates. It is thus important to increase the number of VMS at solar metallicity, and to discover new VMS at higher and lower metallicities (above \zsun\ and below 0.4~\zsun). The galaxies listed above allow probing the range $\sim$0.1 to 1.5~\zsun\ owing to the metallicity gradients that exist from their centers to their outer parts. Table~\ref{tab_param} gathers the sample sizes and metallicity for various degrees of expected progress.

We thus expect to target tens of VMS host candidates. If only a small fraction of these clusters host a few VMS each, this will already multiply the number of individual stars known by a factor of a few, which is a significant improvement over the current population, especially if different metallicities are explored. 

The observational cost should be small since the targets are bright by expected HWO standards (see below).

A preliminary list of targets is as follows:

Individual clusters:
\begin{itemize}
\item NGC5253-5 (3.5 Mpc, 0.4 \zsun)
\item NGC3125-A1 (15 Mpc, 0.4 \zsun)
\item MrK71-A (3.5 Mpc, 0.2 \zsun)
\item II Zw 40 – N (11 Mpc, 0.2 \zsun)
\end{itemize}

Galaxies:
\begin{itemize}
\item M101 (6.4 Mpc, 0.2 to 1 \zsun) $>$ 10 clusters
\item M83 (4.5 Mpc, 0.8 to 1.5 \zsun) $>$ 150 clusters
\item NGC 4449 (4 Mpc, $\sim$0.2 \zsun)
\item NGC 4038/4039 The Antennae (20 Mpc, $\sim$\zsun) $>$ 50 clusters
\item NGC 3310 (17 Mpc, 0.2 to 1 \zsun) $>$ 20 clusters
\end{itemize}

\begin{table*}[!ht]
  \caption{Physical parameters of VMS to be probed for various levels of progress.}
  \label{tab_param}
\smallskip
\begin{center}
{\small
\begin{tabular}{lllll}  
\tableline
\noalign{\smallskip}
Physical Parameter & State of the Art & Incremental Progress & Substancial Progress & Major Progress \\
 & & (Enhancing) & (Enabling) & (Breakthrough)  \\
\noalign{\smallskip}
\tableline
\noalign{\smallskip}
Teff, L, mass, wind parameters & 20 & 40 & 100 & $>$500 \\
Metallicity & 0.4-1.0 \zsun & 0.2 \zsun & 1.5 to 0.1 ~\zsun & $<$0.1~\zsun\\
\noalign{\smallskip}
\tableline\
\end{tabular}
}
\end{center}
\end{table*}

\section{Description of observations}

\subsection{Spatially resolved spectroscopy}

To identify VMS in starburst regions or clusters we need 1) to be able to resolve them spatially and 2) to obtain UV – optical spectroscopy of the resolved components. 

\begin{figure}[ht]
\includegraphics[width=\columnwidth]{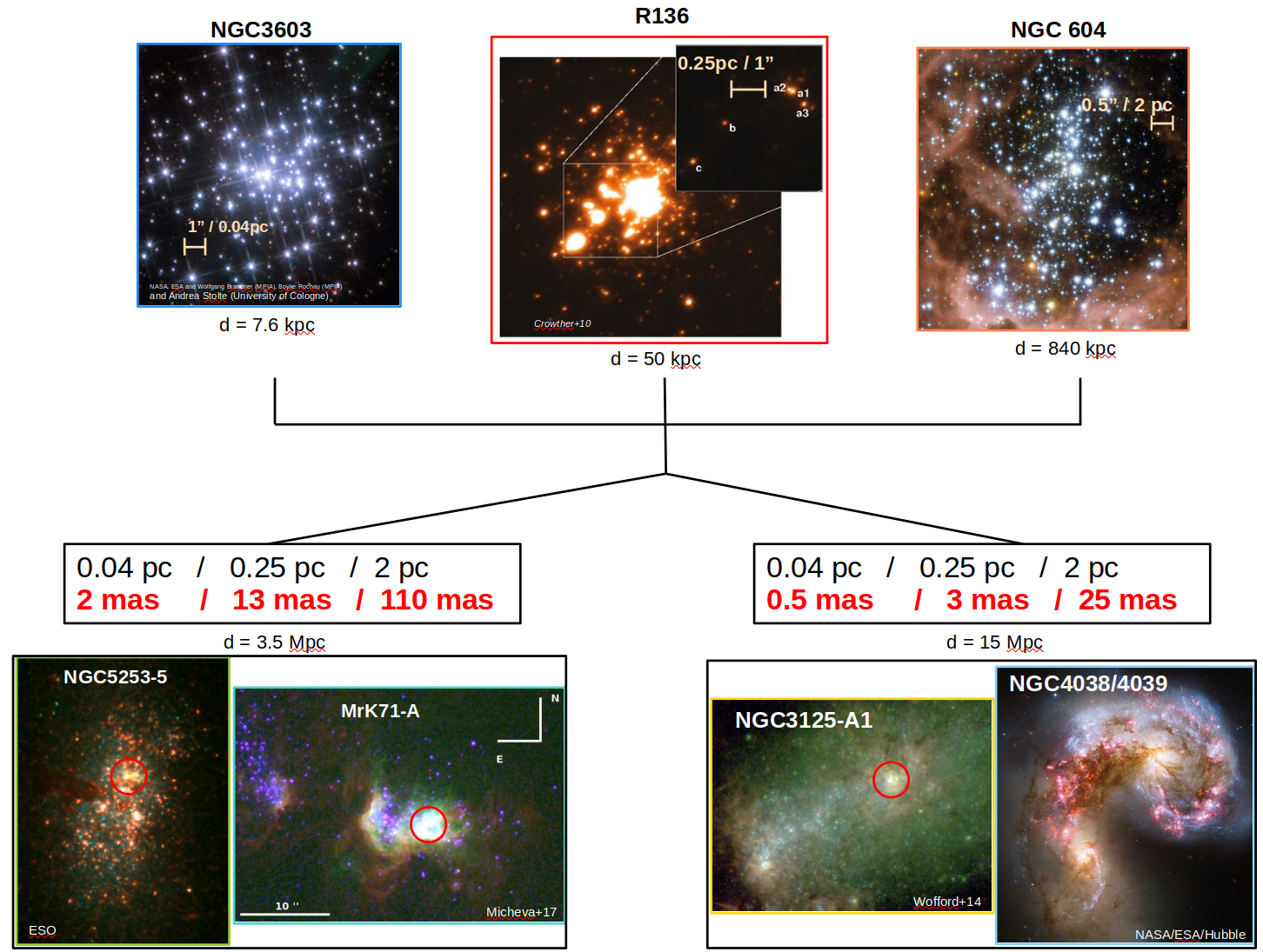}
\caption{\small The three top images show resolved clusters hosting VMS. The size bars indicate typical separation between the most massive components. The bottom panels show images of unresolved clusters likely hosting VMS. Numbers in red are the requested spatial resolution needed to resolve individual VMS in those clusters, assuming their separation is similar to that in the resolved clusters. }
\label{fig_spat_resol}
\end{figure}

To estimate the spatial resolution needed one can use resolved clusters - in which VMS have been observed individually - as templates. This is done in Fig.~\ref{fig_spat_resol}. We take NGC3603, R136 and NGC604 as typical examples since they represent different cases of spatial resolution. From the top sub-panels of Fig.~\ref{fig_spat_resol} we see that the separation between VMS in these clusters ranges from ~0.04 to ~2 pc. These numbers are conservative in the sense that some VMS are separated by less than the adopted typical distances. Unresolved clusters in which the presence of VMS is suspected are shown in the bottom images of Fig. 6. See above for more information on them. Their distances range from 3 to 15 Mpc. Assuming the separation between VMS in these unresolved clusters is the same as that in the resolved clusters, one obtains different constraints on the requested spatial resolution: from 0.5 to 110 mas.

Resolving VMS in a NGC3603-like cluster is thus unfeasible, whatever the distance (the spatial resolution needed being 0.5 to 2 mas). For R136-like clusters, some VMS can be resolved individually at the smallest distances (a few Mpc) but not in the Antennae nor NGC3125-A1 (15 Mpc), unless the UV diffraction limit is reached. VMS that are very close to each other in R136 (i.e. a1, a2 and a3) cannot be resolved in distant clusters. For more extended clusters like in the giant HII region NGC604 VMS are resolved with the current foreseen spatial resolution ($\sim$15 mas).

\begin{figure}[ht]
\includegraphics[width=\columnwidth]{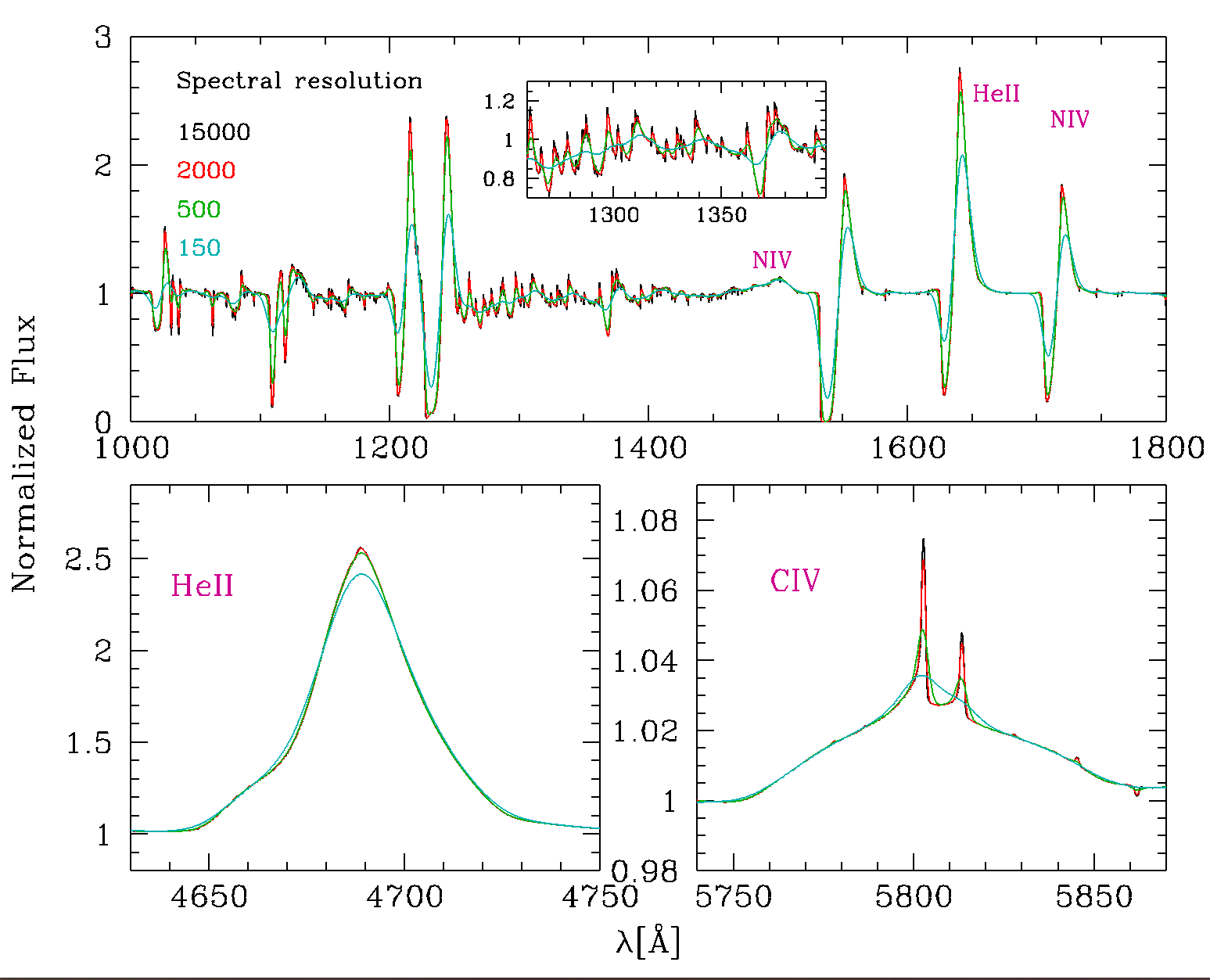}
\caption{\small Typical synthetic spectrum of a VMS in the UV range (top panel) and around \heiiopt\ (bottom left) and \civopt\ (bottom right). Different colors correspond to different spectral resolutions indicated in the top panel. The main lines specific to VMS are labelled in magenta.}
\label{fig_spec_resol}
\end{figure}

The second constraint is for spectroscopy. The identification of VMS and the determination of their physical parameters comes from the UV spectrum between 1200 and 1750~\AA\ and around the \heiiopt\ and \civopt\ features, as described above. Fig.~\ref{fig_spec_resol} shows a typical synthetic spectrum of a VMS with the relevant lines indicated: \heiiuv, \nivuva, \heiiopt, \civopt. The synthetic spectrum is convolved to mimic different spectral resolutions. A minimum of R$\sim$2000 is requested to correctly resolve most features. In particular the doublet in \civopt\ cannot be distinguished from a broad emission at lower spectral resolution, preventing the distinction between VMS and normal WR stars. The iron forests around 1300-1400~\AA\ are also spectrally resolved at R$\sim$2000, enabling metallicity estimates.

VMS are bright objects, with typical absolute magnitudes of -10 in the far UV and -7 in the V band. At the distance of the targets considered in this case (3 to 15 Mpc) their magnitudes are thus expected to be 17 to 21 (UV) and 20 to 24 (V), without considering extinction. Obtaining a SNR of a few tens to correctly detect spectral lines should thus be feasible in a few seconds to a few minutes with the current foreseen sensitivity, which is thus not an issue for the present science case.

\subsection{Observational set-up}

The science case is best conducted with spatially resolved spectroscopy, which can be done with an integral-field spectrograph (IFU). NGC604 is the largest template region in Fig.~\ref{fig_spat_resol}. At a distance of 3.5 (15 Mpc) it can be fully covered in a single pointing with a $6\arcsec \times 6\arcsec$ ($1.5\arcsec \times 1.5\arcsec$) field of view (FoV). The other clusters request smaller FoVs. An IFU with a FoV of $1.5\arcsec \times 1.5\arcsec$ to $3\arcsec \times 3\arcsec$ with the spatial resolution described above and a spectral resolution of R$\sim$2000 would thus be relevant.

Imaging at the same spatial resolution can be performed with the planned UV-optical imager and should deliver photometry required to constrain luminosity. It is likely that ground-based ELTs will also be able to provide photometry of resolved sources, but at longer wavelength (in particular in the near-IR). This will be useful for luminosity constraints so HWO imaging is not fully mandatory, contrary to spatially resolved UV-optical spectroscopy. But combining UV-optical with near-IR photometry will set better constraints on the SEDs of VMS, and thus will lead to more robust luminosity and mass estimates. 

Table~\ref{tab_require} summarizes the main observational requirements to obtain spatially rtesolved spectroscopy of VMS in different environments.

\subsection{The need for HWO}

Studying VMSs separately from their host clusters is a unique science case to HWO. The ELTs (e.g. the finest scale of ELT-MICADO and ELT-HARMONI) could resolve the sources in the NIR. However, adaptive optics is not functional at the optical range and UV is not accessible to ground-based ELTs. As of today the only diagnostic features of VMS are located in the UV, and to a lesser extent in the optical range. Hence the unambiguous identification and characterization of VMS can only be achieved with UV spectroscopy. IFU spectroscopy is the most efficient way to observe VMSs, separate them from their close neighbours and simultaneously capture other massive stars in starburst environments. This way, the upper initial mass function can be uniquely characterized.

\begin{table*}[!ht]
  \caption{Observation requirements to spectroscopically resolve individual VMS in different environments}
  \label{tab_require}
\smallskip
\begin{center}
{\small
\begin{tabular}{lllll} 
\tableline
\noalign{\smallskip}
Observarion Requirement & State of the Art & Incremental Progress & Substancial Progress & Major Progress \\
 & & (Enhancing) & (Enabling) & (Breakthrough)  \\
\noalign{\smallskip}
\tableline
\noalign{\smallskip}
UV IFU & None &  & Optical diffraction limited & UV diffraction limited \\
VMS resolved spectroscopy & UV-NIR & NIR & optical+UV & optical+UV \\
 & d$<$0.8Mpc & 0.8$<$d$<$ $\sim$3.5 Mpc (ELT) & 0.8$<$d$<$ $\sim$3.5 Mpc & 0.8$<$d$<$ $\sim$15 Mpc \\
 & 20 stars & & & \\
Wavelength Range & none for individual & $>$6000~\AA\ (ELT) & 1200-6000~\AA & 1000-7000~\AA \\
 & stars at d$>$0.8Mpc &  & & \\
Spectral resolution & none for individual & R$>$500 & R$>$2000 & R$>$5000 \\
 & stars at d$>$0.8Mpc &  & & \\
Spatial resolution & 50 mas & 4 mas NIR (ELT) & 15 mas UV & 5mas UV \\
\noalign{\smallskip}
\tableline\
\end{tabular}
}
\end{center}
\end{table*}

\vspace{0.5cm}

{\bf Acknowledgements.} We thank the HWO teams for help, support, and enthusiasm in the preparation of this science case.

\bibliography{VMS_hwo.bib}

\end{document}